\documentclass[aps,prl,twocolumn,superscriptaddress]{revtex4}

\usepackage{amsfonts,epsfig}
\usepackage{latexsym}
\usepackage{epsfig}
\usepackage{amsmath}
\usepackage{amssymb}
\usepackage{amsfonts}
\usepackage{amsthm}
\usepackage{mathrsfs}
\usepackage{natbib}
\usepackage{color,verbatim,graphics}
\DeclareMathAlphabet{\mathrsfs}{U}{rsfs}{m}{n}
\DeclareMathAlphabet{\mathpzc}{OT1}{pzc}{m}{it}
\DeclareMathAlphabet{\matheus}{U}{eus}{m}{n}
\DeclareMathAlphabet{\mathbbold}{U}{bbold}{m}{n}

\newcommand{\ba}{\begin{eqnarray}}
\newcommand{\ea}{\end{eqnarray}}
\newcommand{\ban}{\begin{eqnarray*}}
\newcommand{\ean}{\end{eqnarray*}}

\newcommand{\ket}[1]{|#1\rangle}
\newcommand{\bra}[1]{\langle#1|}

\newcommand{\eg}{{\it{e.g.~}}}

\begin{document}

\title{Large violation of Bell inequalities using both particle and wave measurements}

\author{Daniel Cavalcanti}
\email{dcavalcanti@gmail.com}
\affiliation{Centre for Quantum Technologies, National University of Singapore, 3 Science drive 2, Singapore 117543}%
\author{Nicolas Brunner}
\affiliation{H.H. Wills Physics Laboratory, University of Bristol, Tyndall Avenue, Bristol, BS8 1TL, United Kingdom}
\author{Paul Skrzypczyk}
\affiliation{H.H. Wills Physics Laboratory, University of Bristol, Tyndall Avenue, Bristol, BS8 1TL, United Kingdom}
\author{Alejo Salles}
\affiliation{Niels Bohr Institute, Blegdamsvej 17, 2100 Copenhagen, Denmark}
\author{Valerio Scarani}
\affiliation{Centre for Quantum Technologies, National University of Singapore, 3 Science drive 2, Singapore 117543}
\affiliation{Department of Physics, National University of Singapore, 2 Science Drive 3, Singapore 117542}



\begin{abstract}
When separated measurements on entangled quantum systems are performed, the theory predicts correlations that cannot be explained by any classical mechanism: communication is excluded because the signal should travel faster than light; pre-established agreement is excluded because \textit{Bell inequalities} are violated. All optical demonstrations of such violations have involved discrete degrees of freedom and are plagued by the detection-efficiency loophole. A promising alternative is to use continuous variables combined with highly efficient homodyne measurements. However, all the schemes proposed so far use states or measurements that are extremely difficult to achieve, or produce very weak violations. We present a simple method to generate large violations for feasible states using both photon counting and homodyne detections. The present scheme can also be used to obtain nonlocality from easy-to-prepare Gaussian states (\eg two-mode squeezed state).
\end{abstract}

\maketitle



The violation of Bell inequalities has played a crucial role in the foundations of quantum physics, since it provides a testable criterion to rule out classical mechanisms as the origin of quantum correlations  \cite{bell64}. Moreover, it is also an important test for future applications, since it provides device-independent assessment of the performance of some quantum tasks like key distribution \cite{diqkd} or randomness generation \cite{dirandom}.

In experiments, violations have been demonstrated so far only for discrete-outcome measurements \cite{genovese}. The countless optical realizations have used several encodings, the most frequent ones being polarization \cite{aspect,weihs} or time-bins \cite{tittel}. Light can easily be sent at large distances, so the locality loophole can be closed; but the detection loophole \cite{pearle} remains open due to the joint effect of losses (both in the coupling between the source and the optical link, and in the link itself) and of limited efficiency of the photon counters. When energy levels of ions and atoms are used, fluorescence measurements are very efficient but slow: the detection loophole can be closed \cite{rowe,dirandom}, but it is practically impossible to think of separating these systems far enough to close the locality loophole. Entanglement swapping between light and atoms was proposed several years ago in order to combine the best of both worlds \cite{simon}, but its full implementation has yet to be reported \cite{rosenfeld}.

Another path towards a loophole-free Bell test consists in using only light, but measuring rather continuous degrees of freedom, exploiting the high efficiency of homodyne measurements \cite{reid}. However, this path has proved harder than expected: no experimental violation of Bell inequalities (let alone loophole-free ones) involving homodyne measurements has been reported to date.
One of the main problems is that for the simplest states that can be produced (having positive, usually Gaussian, Wigner functions),
homodyne measurements produce statistics that do not violate any Bell inequality. Some theoretical schemes have shown however that violations are indeed possible, however they require either measurements \cite{bana,stobi} or states \cite{munro,wenger,ecava,acin} that are practically unfeasible. Only in 2004 a
proposal was put forth~\cite{grangier1,nhacar}, in which homodyne measurements on a feasible state, followed by suitable data processing, lead to a violation $S\approx 2.046$ of the Clauser-Horne-Shimony-Holt (CHSH) inequality $S\leq 2$ \cite{chsh}. Such a small violation, however, is  hardly observable in the presence of imperfections, and has indeed not yet been achieved experimentally.

The main goal of the present paper is to demonstrate that large violations of Bell inequalities can indeed be achieved with feasible setups involving homodyne measurements. We study schemes in which both Alice and Bob alternate between photon counting and homodyne measurements, then locally post-process their data to extract bits and check the CHSH inequality.
We show that a significant violation $S\approx 2.25$ can be achieved by the state
\ba\ket{\Psi_2}=\frac{\ket{2}_A\ket{0}_B+\ket{0}_A\ket{2}_B}{\sqrt{2}},
\label{noonstate}\ea
where again $\ket{0}$ and $\ket{2}$ refer again to states of well defined photon-number. This state can be created by having two heralded single photons from down-conversion sources bunch on a beam-splitter, in a Hong-Ou-Mandel setup \cite{hom}.

Our scheme was motivated by a recent result by Ji and coworkers in the tentative of finding Bell tests for easy-to-prepare quantum states \cite{koreans}. However, the inequalities they used are not Bell inequalities in the most general sense, since they rule out only a particular class of local models. Thus they cannot be used for any device-independent assessment---as required for demonstrating nonlocality---since they can be violated by a local model \cite{comment}.



\begin{figure}[ht]
\begin{tabular}{cc}
\includegraphics[scale=0.3]{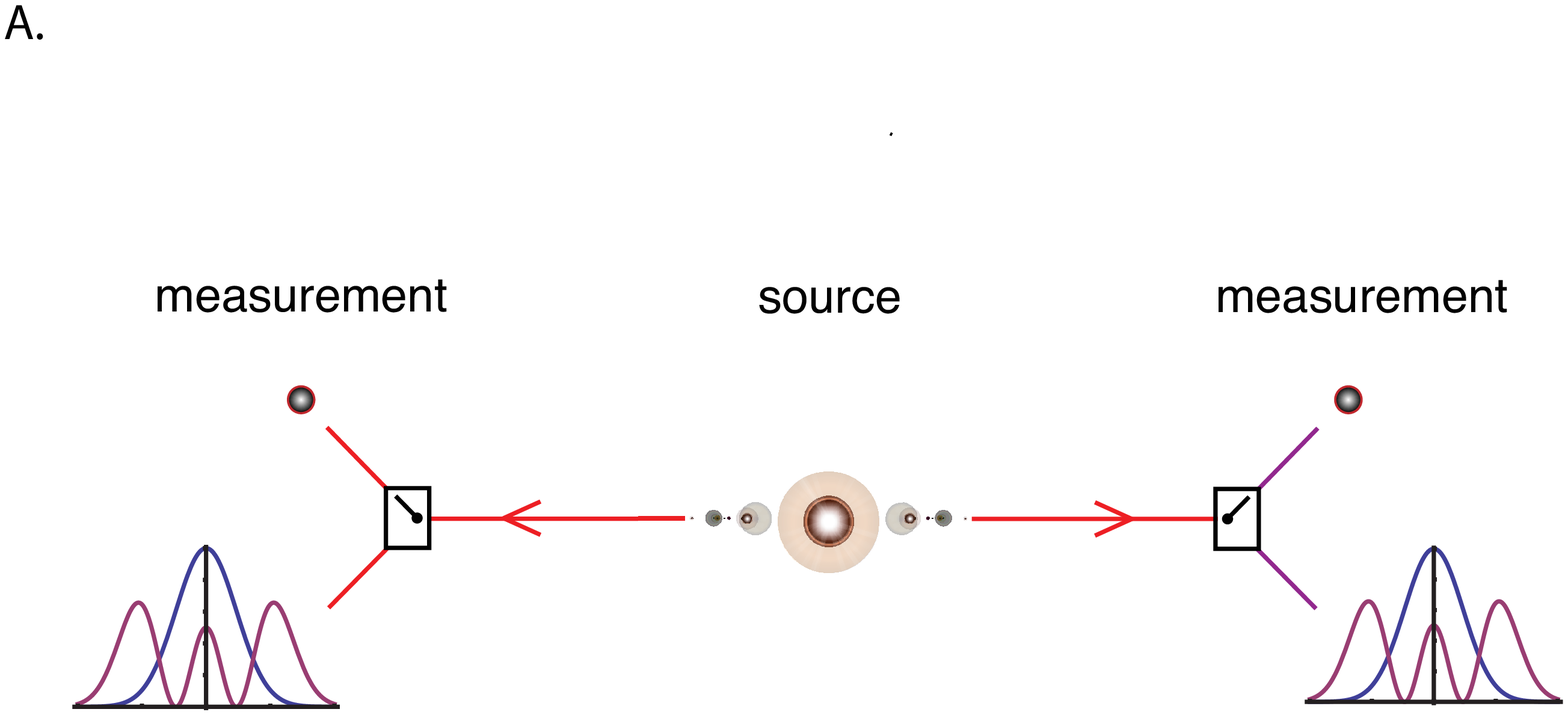}\\
\includegraphics[scale=0.3]{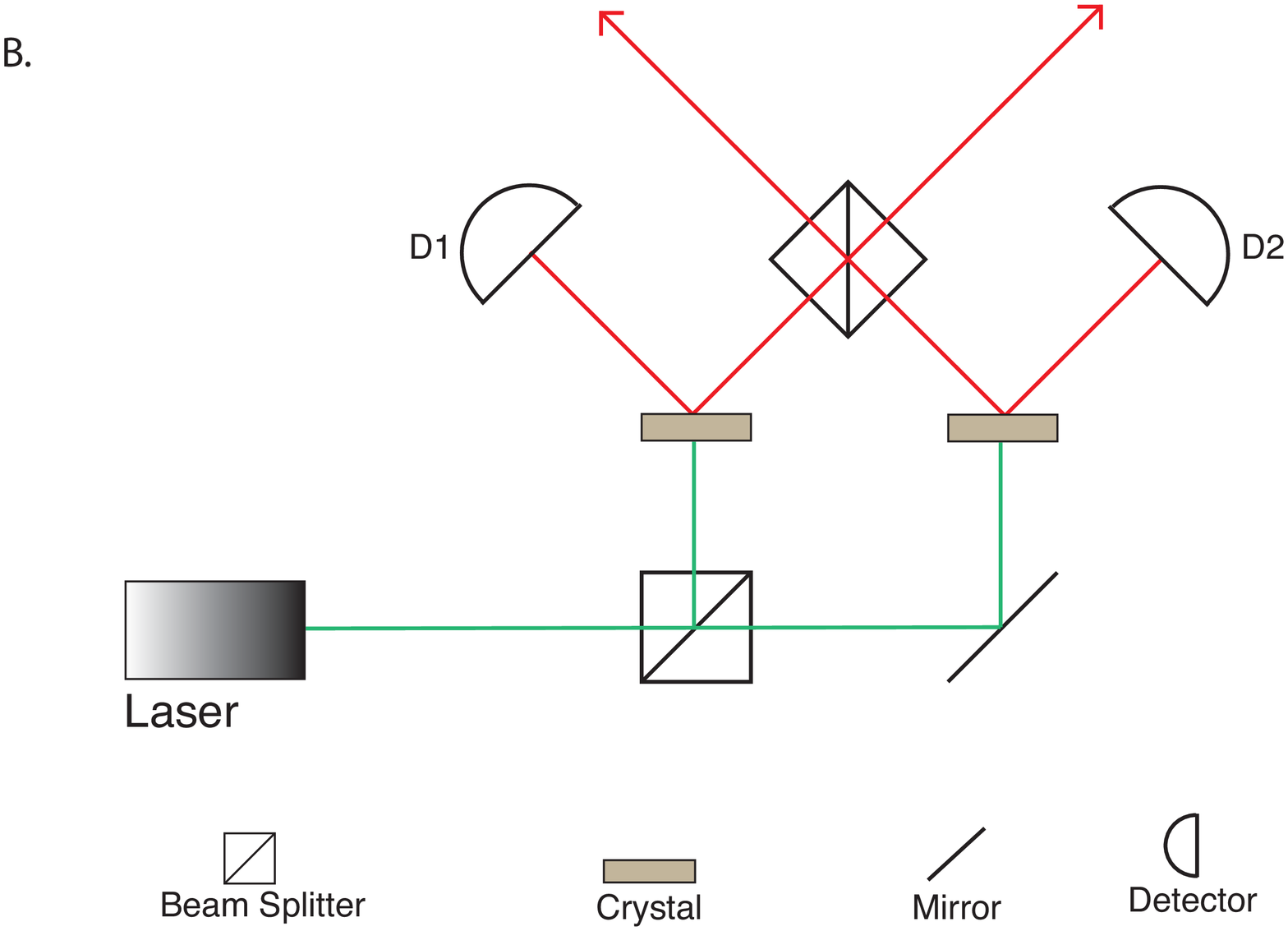}
\end{tabular}
	\caption{{(Color online) Sketch of the setup.} {\bf A.} A source sends a photonic entangled state to two space-like separated locations. In these locations each subsystem is subjected to one of two measurements: number of photons (photon counting) or quadrature (homodyning) measurements.  In this way both ``wave'' and ``particle'' characteristics of the systems are tested.  {\bf B.} The state $(\ket{0}\ket{2}+\ket{2}\ket{0})/\sqrt{2}$ violates the CHSH Bell inequality in the previous scenario and can be created as follows: two pairs of photons are created in different non-linear crystals by parametric down conversion. The detection of one photon of each pair at detectors D1 and D2 heralds the presence of the other two photons, which are sent to a beam splitter. The Hong-Ou-Mandel interference in the beam splitter makes the photons bunch, resulting in the desired two-photon state.}\label{fignice}
\end{figure}

\textbf{Ideal case.--}
The setup under study is sketched in Fig.~\ref{fignice}. Alice and Bob can perform two measurements each: one is the photon number $N$; the other is the $X$ quadrature. The measurement results are then processed to obtain bits $a,b\in\{-1,+1\}$, where $a$ and $b$ label Alice and Bob's outcomes respectively. We describe these binning procedures for the case of Alice, those of Bob are identical. When measuring $N$, Alice sets $a=+1$ if the result is $N>0$ and $a=-1$ if the result is $N=0$: this binning is simply the direct outcome of a perfect threshold detector. As for the $X$ measurement, Alice divides the real axis in two disjoint regions and sets $a=+1$ if $x\in {\cal A}^+$ and $a=-1$ if $x\in {\cal A}^-=\mathbb{R}\setminus {\cal A}^+$. These sets can still be quite complicated in general; here it will be sufficient to consider very simple sets, namely ${\cal A}^+={\cal B}^+=[-z,z]$, where $z$ remains to be chosen.


\begin{figure}[ht]
	\includegraphics[scale=0.5]{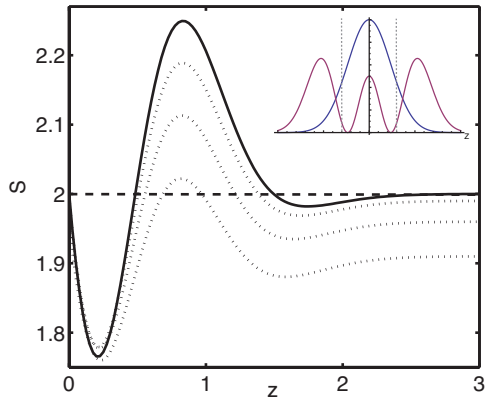}
	\caption{{(Color online) Value of the CHSH expression S as a function of the parameter $z$ for ideal detectors.} The full line is for the state $\ket{\Psi_2}$ given in (\ref{noonstate}). The dashed lines are for $\rho$ given in (\ref{eqrho}), describing a lossy line with transmisivity $t=0.95$, $0.9$ and $0.85$ from top to bottom respectively. Inset: the density functions $|\phi_0(x)|^2$ (full blue line) and $|\phi_2(x)|^2$ (dashed purple line), with the choice $z\approx 0.83$ (dotted vertical lines) for the maximal violation $S\approx 2.25$; notice that this value of $z$ allows one to discriminate the density functions with high probability. This is important to attain a high violation of CHSH since it allows to maximize the correlations between the $X$ and $N$ measurements. }\label{figst}
\end{figure}

Using these measurements, we focus on the CHSH inequality, which reads
\ba
S&=&E_{XX}+E_{XN}+E_{NX}-E_{NN}\leq2,\label{chshXN}
\ea
where $E_{jk}=P(a=b|jk)-P(a\neq b|jk)$ is the expectation value of the measurements $j$ and $k$ after the binning.
Now we are going to show that this inequality can be violated by measuring the state \eqref{noonstate}. The statistics of the four pairs of measurements are easy to write down. In fact, when both Alice and Bob measure $N$, their bits are always different, hence $E_{NN}=-1$. When Alice measures $N$ and Bob measures $X$: if $a=+1$, Bob's state is $\ket{0}$, whence his measurement of $X$ is described by the density function $|\langle x|0\rangle|^2=|\phi_0(x)|^2$ where $\phi_0(x)=\frac{1}{\pi^{1/4}}e^{-x^2/2}$; similarly, if $a=-1$, Bob's statistics are described by the density $|\langle x|2\rangle|^2=|\phi_2(x)|^2$ where $\phi_2(x)=\frac{1}{(4\pi)^{1/4}}(2x^2-1)e^{-x^2/2}$. The case when Alice measures X and Bob measures N is symmetric. Finally, when both Alice and Bob measure $X$, their statistics are described by $|\langle x_A,x_B|\Psi_2\rangle|^2=|\Psi_2(x_A,x_B)|^2$ where $\Psi_2(x_A,x_B)$ is obtained by replacing the state $\ket{k}$ with $\phi_k(x)$ in (\ref{noonstate}). All in all, the probabilities are given by the following expressions:
\ba
\begin{array}{lcl}
P(a,b|NN)&=&(1-ab)/4\,;\\
P(a,b|XN)&=&\frac{1}{2}\int_{{\cal A}^a}dx|\phi_{m(b)}(x)|^2\,;\\
P(a,b|NX)&=&\frac{1}{2}\int_{{\cal B}^b}dx|\phi_{m(a)}(x)|^2\,;\\
P(a,b|XX)&=&\int_{{\cal A}^a} dx\int_{{\cal B}^b}dy |\Psi_2(x,y)|^2,\end{array}\label{qprobs}
\ea where $m(+1)=0$ and $m(-1)=2$.
Substituting these statistics into (\ref{chshXN}), one obtains a value of $S$ for any choice of $z$. The maximal violation of the CHSH inequality is $S\approx 2.25$ for $z\approx 0.83$ (see Fig.~\ref{figst}).


\textbf{Non-ideal case.--}
So far, we have proved that an ideal realization of the state (\ref{noonstate}) would lead to a large violation of CHSH for ideal detectors. Let us now introduce two deviations from the ideal case and study the robustness of the result (for simplicity, all the parameters below are supposed to be the same for Alice and Bob).

First, we introduce the \textit{transmission} $t$ of the optical paths between the sources and the detectors. This parameter includes the coupling from the source into the transmitted mode and the subsequent possible losses in the channel. The ideal state $\ket{\Psi_2}$ reaches the detectors with probability $t^2$. With probability $2t(1-t)$, one of the two photons is lost. In this case, the state at the detector becomes $\rho_1=\frac{1}{2}(\ket{10}\bra{10}+\ket{01}\bra{01})$, because the photon lost in the environment would identify the path. Finally, with probability $(1-t)^2$, both photons are lost and the state at the detector is just $\ket{00}$. The final state measured is therefore
\ba
\rho= t^2\ket{\Psi_2}\bra{\Psi_2}\,+\,2t(1-t)\,\rho_1\,+\,(1-t)^2\ket{00}\bra{00}.
\label{eqrho}
\ea

\begin{figure}[ht]
\includegraphics[scale=0.5]{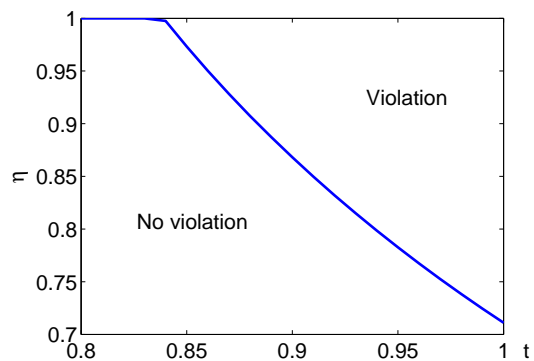}
\caption{{(Color online) Quantum efficiency of the threshold detectors ($\eta$) vs. transmission of the optical links ($t$) for violation of the CHSH inequality.} The curve supposes that, for each $t$, the optimal choice of $z$ for the binning is made. If there are no losses in the line, the detector efficiency can be as low as $\eta\approx 71.1\%$; conversely, for perfectly efficient detectors, one can tolerate a transmission $t\approx 84\%$.}\label{figetat}
\end{figure}

Second, while keeping the measurement $X$ fully efficient, we attribute a \textit{quantum efficiency} $\eta<1$ to the threshold detector used to perform the measurement $N$. We stress that no post-selection will be performed on the data: each event in which the threshold detector does not fire will be counted as $a=-1$, respectively $b=-1$. The final result is shown in Fig.~\ref{figetat} (see also Methods). Our scheme is more sensitive to losses on the line than to losses on the threshold detector: this was expected, since the former affect both measurements while the latter affect only the $N$ measurements. For a transmission of $t=90\%$, a detection efficiency of $\eta \approx 86\%$ can be tolerated. Though these are demanding features, they are within reach of current technology \cite{pittman,rosenberg}. These numbers are also comparable to the most favorable feasible schemes known to date for discrete variables, where the figure of merit is $\eta t$ \cite{vpb}. In contrast, here the losses correspond to the imperfections
of the state (since they act on the same degree of freedom as the
measurements) while for discrete variables the imperfections of the state
are an additional problem.


\textbf{Experimental considerations.--}
Let us make some considerations about the experimental implementation of our scheme. Homodyne measurements require a sufficiently long coherence time of the signal. So, if the state $\ket{\Psi_2}$ is implemented using down-conversion sources as we propose, the bandwidth of the down-converted photons must be narrow enough. Fortunately, the Hong-Ou-Mandel effect between photons coming from different crystals has been demonstrated using continuous pumping \cite{halder}; thus pulsed pumping is not a hidden requirement (notice that this fact has another positive consequence: the four-photon processes in down-conversion can indeed be neglected). Also homodyne measurements on one and two-photon states coming from down-conversion have already been reported \cite{lvo}. It seems therefore that the experiment is feasible with current technology, though certainly challenging.

\textbf{Other quantum states.--}
The combination of counting and homodyne measurements can be applied to many more scenarios. A natural question is whether \textit{other states}, among those that are feasible in laboratories today, violate the CHSH inequality. It turns out that the two-mode squeezed state
\ba
\ket{\psi}&=&\sqrt{1-\lambda^2}\sum_{n}\lambda^n \ket{n}\ket{n}
\label{TMSS}
\ea
violates a version of CHSH for some values of $\lambda$, provided (say) Bob's homodyne measurements is in the complementary quadrature $P$.
Although the violation found is small ($S\approx 2.05$ for $\lambda\approx 0.83$ and $z\approx0.86$), it is remakable, since this state is Gaussian and easily produceable in the lab. Note also that the amount of violation is similar to the best value previously reported with a feasible state \cite{grangier1,nhacar}. The latter, however, used a more complicated state, obtained from \eqref{TMSS} by photon subtraction in each arm.
We could not find any violation for the states (here unnormalized) $\ket{1}\ket{0}+\ket{0}\ket{1}$ \cite{0110} and $\ket{\alpha}\ket{-\alpha}+\ket{-\alpha}\ket{\alpha}$ ($\ket{\alpha}$ being a coherent state of amplitude $\alpha$) \cite{gran}.

\textbf{Discussion.--}
The use of efficient homodyne measurements and photonic continuous-degrees of freedom in Bell tests has triggered much attention in the past years. Although this appears as an interesting path towards a loophole-free nonlocality test, no result so far had indicated that this method could actually work in practice. All the  results reported previously suffered from using impractical quantum states and measurements, or achieved very small violations. Our main goal was to overcome these problems and present a feasible scheme to observe a large violations of Bell inequalities with continuous-variable measurements. The key element was to combine both photon counting and homodyne measurements in the same Bell test.

Although the implementation of our explicit scheme is still challenging, we believe that our method opens up new possibilities for designing a loophole-free Bell test. From the theoretical point of view, considering other quantum states and/or more sophisticated Bell inequalities could lead to larger and more robust violations. From the experimental point of view, simplifying the creation of the states described here and the progress towards the experimental considerations we discussed are certainly fruitful ways to research on.

Finally, we believe that a proof-of-principle experiment in which the experimental data is post-processed in order to take into account the inefficiencies in the experiment (similarly to the fair-sampling assumption in the discrete case) is interesting in its own right. Such an experimental demonstration would reinforce the usefulness of homodyne measurements in Bell tests and could be realized with current technology.

\textbf{Appendix.--}
In order to study the effect of the limited efficiency $\eta$, we rewrite the CHSH inequality in the Clauser-Horne form \cite{CH}, which is equivalent for no-signaling distributions:
\ba\label{CHineq}
&-&p(a_X=+)-p(b_X=+)+p(++|XX)\\&+&p(++|NX)+p(++|XN)-p(++|NN)\leq0.\nonumber
\ea
Here, $p$ describes the \textit{observed} statistics. Now, $p(++|NN)=0$ because one of the modes is always empty (we are neglecting spurious counts here). The first line can be re-written as $p(--|XX)-1$ and there is no effect of $\eta$, so one just has to compute this quantity for $\rho$ along the same lines as we did for $\ket{\Psi_2}$ above. Finally, consider $p(++|NX)$, the case for $p(++|XN)$ being symmetric. If the state is $\ket{\Psi_2}$, one has $p(++|NX,\Psi_2)=[1-(1-\eta)^2]\,P(++|NX)$ where $P(++|NX)$ is given in (\ref{qprobs}), because there are two photons reaching the detector. If the state is $\rho_1$, one has $p(++|NX,\rho_1)=\eta\,P(++|NX)$: indeed, Alice finds $a_N=+1$ with probability $\frac{\eta}{2}$ and prepares the state $\phi_0(x)$ on Bob's side. When the state is $\ket{00}$, Alice never finds $a_N=+1$. All in all, $p(++|NX)=t\eta(2-t\eta)\,P(++|NX)$. Thus the condition for \eqref{CHineq} to be violated becomes
\ba
t\eta&\geq&1-\sqrt{1-\frac{1-p(--|XX)}{P(++|NX)+P(++|XN)}}\,.
\ea Note that $t$ enters in the r.h.s. of this equation through $p(--|XX)$ evaluated for $\rho$. So, contrary to the schemes using discrete variables, the effects of $t$ and $\eta$ are not identical. Ultimately, one has to resort to numerical evaluation to find the best value of $z$ for each case.

AS, NB and PS acknowledge hospitality from the National University of Singapore. DC thanks A. Ac\'in and A. Ferraro for many discussions about non-locality in continuous-variable systems along the years.
We thank C. Kurtsiefer and D. Tasca for useful discussions. This work was supported by the National Research Foundation and the Ministry of Education, Singapore, the UK EPSRC, EU STREP COQUIT under FET-Open grant number 233747, and QESSENCE.

\end{document}